\documentclass{new_tlp}
\usepackage{mathptmx}

\usepackage{amsmath}
\usepackage{amssymb}
\usepackage[e]{esvect}
\usepackage{graphicx}
\usepackage{multicol}
\usepackage{parcolumns}
\usepackage{amsfonts}
\usepackage{fancyvrb}
\usepackage{xcolor}

\newcommand{\leanparagraph}[1]{\smallskip\noindent\textbf{#1}. }

\newcommand{\ext}[3]{\ensuremath{\amp{#1}[#2](#3)}}

\DeclareMathOperator{\naf}{not}

\newcommand{\extfun}[1]{\ensuremath{f_{\text{\sl\&}#1}}}

\newcommand{\extsem}[4]{\ensuremath{f_{\text{\sl\&}#1}(#2,#3,#4)}}

\newcommand{\amp}[1]{\ensuremath{\text{\textsl{{\&}}}\!\mathit{#1}}}

\newcommand\hex{{\sc hex}}

\newcommand{\BK}{BK}
\newcommand{\milp}{\mathcal{M}}

\newtheorem{theorem}{Theorem}

\newtheorem{definition}{Definition}
\newtheorem{example}{Example}

\newcommand\bcmdtab{\noindent\bgroup\tabcolsep=0pt%
  \begin{tabular}{@{}p{10pc}@{}p{20pc}@{}}}
\newcommand\ecmdtab{\end{tabular}\egroup}

  \title[Exploiting ASP with
External Sources for Meta-Interpretive Learning]
        {Exploiting Answer Set Programming with\\
External Sources for Meta-Interpretive Learning}

 \author[Tobias Kaminski, Thomas Eiter and Katsumi Inoue]
         {
	 TOBIAS KAMINSKI, THOMAS EITER\\
         Technical University of Vienna (TU Wien), Vienna, Austria\\
         \email{\{kaminski,eiter\}@kr.tuwien.ac.at}
         \and KATSUMI INOUE\\
         National Institute of Informatics, Tokyo, Japan\\
         \email{inoue@nii.ac.jp}}

\jdate{March 2003}
\pubyear{2003}
\pagerange{\pageref{firstpage}--\pageref{lastpage}}
\doi{S1471068401001193}

\begin{document}

\label{firstpage}

\maketitle

  \begin{abstract}
Meta-Interpretive Learning (MIL) learns logic programs from examples
by instantiating meta-rules, which is implemented by the Metagol system based on Prolog.
Viewing MIL-problems as
combinatorial search problems, they can alternatively be solved by
employing Answer Set Programming (ASP), which may result in performance
gains as a result of efficient conflict propagation.
However, a
straightforward ASP-encoding of MIL results in a huge search space due
to a lack of procedural bias and the need for grounding. To address
these challenging issues, we encode MIL in the HEX-formalism, which is
an extension of ASP that allows us to outsource the background
knowledge, and we restrict the search space to compensate for a procedural
bias in ASP. This way, the import of constants from the background knowledge
can for a given type of meta-rules be limited to relevant
ones. Moreover, by abstracting from term manipulations in the encoding
and by exploiting the HEX interface mechanism, the import of such
constants can be entirely avoided in order to mitigate the grounding
bottleneck. An experimental evaluation shows promising results.\\[0.1cm]
\emph{Note:} This paper is under consideration for acceptance in TPLP.
  \end{abstract}



\section{Introduction}
\label{intro}
Recently, \emph{Meta-Interpretive Learning} (\emph{MIL}) \cite{MuggletonLT15} has attracted a lot of attention in the area of \emph{Inductive Logic Programming} (\emph{ILP}).
The approach learns definite logic programs from positive and negative
examples given some background knowledge by instantiating so-called
\emph{meta-rules}. 
The latter can be viewed as templates specifying the shapes of rules that may be used in the induced program.
The formalism is very powerful as it enables \emph{predicate
  invention}, i.e.\ 
to 
use new predicate symbols in the
induced program, and supports learning of recursive programs, while the
hypothesis space can be constrained effectively 
by using meta-rules.

MIL has been implemented in the \emph{Metagol} system \cite{metagol}, which is based on a classical \emph{Prolog} meta-interpreter.
The system is very efficient by exploiting the query-driven procedure
of Prolog 
to guide the instantiation of meta-rules
in a specific order.
In contrast (and complementary) to a common \emph{declarative bias} in
ILP which constrains the hypothesis space, this constitutes  
a {\em procedural bias}\/ that may affect efficiency (or even termination).

While traditionally most ILP systems 
are based on Prolog,
the advantages of 
\emph{Answer Set Programming} (\emph{ASP}) \cite{gelf-lifs-91}
for ILP  
were recognized and several ASP-based systems have been developed, e.g.\
\cite{Otero01,Ray09,LawRB14}.
Some benign features of ASP are its pure declarativity,  
which allows to modularly restrict the search space by adding
rules and constraints to an encoding without risking non-termination,
and that enumeration of solutions is easy.
Furthermore, the efficiency and optimization techniques of modern ASP solvers
such as \textsc{Clasp} \cite{gks2012-aij}, which supports conflict propagation and learning,
can be exploited.
\citeN{MuggletonLPT14} already considered an ASP-version of Metagol,
which used only one specific meta-rule and was tailored to 
inducing grammars. The authors observed that ASP can have
an advantage for MIL over Prolog due to effective pruning, but that it
performs worse when the background knowledge is more extensive or only
few constraints are present.

Implementing general MIL by ASP comes with its own
challenges; and
solving MIL-problems efficiently by utilizing a
straightforward ASP encoding turns out to be infeasible in many cases.
The first challenge 
is 
the large search space 
as a result of an unguided search due 
to a lack of procedural bias. 
Consequently,  
the search space must be carefully restricted in an encoding in order to avoid many
irrelevant instantiations of meta-rules.  The second and more severe
challenge concerns the \emph{grounding bottleneck} of ASP:
in contrast to Prolog, where only relevant terms are
taken into account by \emph{unification}, all terms
that possibly occur in a derivation from the background knowledge 
must be considered in a grounding step.
Finally, a third challenge
are recursive manipulations of structured objects, such as 
strings or lists, that are common
for defining background knowledge in Metagol and easy to realize 
in Prolog, but 
are less supported in ASP.

In this paper, we 
meet the mentioned challenges for a class of MIL-problems that is
widely 
encountered in practice, by developing a MIL-encoding 
in \emph{ASP with external sources},
specifically in the
\hex-formalism
\cite{EiterFIKRS16}.
\hex-programs
extend ASP 
with a bidirectional
information exchange between a
program and arbitrary external sources of computation via special
\emph{external atoms}, which may introduce new constants into a
program by so-called \emph{value invention}.

After introducing the necessary background on MIL and \hex-programs in
Section~\ref{sec:prelim}, we proceed to present our contributions as
follows:

\begin{itemize}
\item We introduce in Section~\ref{sec:enc} 
our novel MIL approach based on \hex-programs for general
MIL-problems. In the first  encoding, $\Pi(\mathcal{M})$, we
restrict the search space by interleaving derivations 
at the object level and 
the meta level such that new instantiations of meta-rules can
be generated based on pieces of information that 
are already derived wrt.\ partial hypotheses of rules.
Furthermore, we outsource the background knowledge and access it by
means of external atoms, which enables the manipulation of complex
objects such as strings or lists.  

\item We then define the class of {\em forward-chained}\/
  MIL-problems, for which the grounding can be restricted. Informally,
  in such
   problems the elements $x,y$ in the binary head
  $p(x,y)$ of a rule must be connected via a path $p_1(x_1,x_2)$,
  $p_2(x_2,x_3)$, \ldots, $p_k(x_k,x_{k+1})$ in the body, where $x=x_1$ and
  $x_{k+1}=y$. This allows us to guard the import of new terms from
  the background knowledge in a second encoding, $\Pi_f(\mathcal{M})$,
  by using already imported terms in an inductive manner.
  
\item A large number of constants may still be imported from the
background knowledge.
We thus develop in Section~\ref{sec:abst} a
  technique to abstract from object-level terms in a third encoding, $\Pi_{sa}(\mathcal{M})$,
  by externally computing sequences of background knowledge atoms that
  derive all positive examples, and by checking non-derivability of negative
  examples 
  with an external constraint.
  
\item In Section~\ref{sec:eval}, we present results of an empirical
  evaluation
based on known benchmark problems;
they provide evidence
for the potential of using a \hex-based approach for MIL.
\end{itemize}

While our encoding is inspired by the implementation presented in
\cite{MuggletonLPT14}, to the best of our knowledge, a general
implementation of MIL using ASP has not been considered in the
literature so far, and neither strategies 
to compensate for the missing
procedural bias nor
to mitigate grounding issues have been
investigated. Despite the use of the \hex\ formalism, our
results may be applied to other ASP formalisms and approaches as well.
Proof sketches and details on the benchmark encodings used in Section~\ref{sec:eval} can be found
 in the appendix.

\section{Background}
\label{sec:prelim}
We assume a finite set $\mathcal{P}$ of predicate symbols,
a finite set  $\mathcal{C}$ of constant symbols,
and disjoint sets $\mathcal{X}_1$ and $\mathcal{X}_2$
of first-order and higher-order variables, resp., not overlapping with $\mathcal{P}$ and $\mathcal{C}$.
An atom $a$ of the form $p(t_1,...,t_n)$, where $t_i \in \mathcal{C}
\cup \mathcal{X}_1$ for $1 \leq i \leq n$, is \emph{first-order} if $p
\in \mathcal{P}$ and \emph{higher-order} if $p \in \mathcal{X}_2$;
its \emph{arity} is $n$.
A \emph{ground atom} is a first-order atom where $t_i \in \mathcal{C}$ for all $1 \leq i \leq n$. We represent \emph{interpretations} over the Herbrand base by sets of ground  atoms, and an interpretation $I$ \emph{models} a ground atom $a$, denoted $I \models a$, if $a \in I$.

A (\emph{disjunctive}) \emph{logic program} $P$ is a set of rules of the form
\begin{equation}
  a_1\lor\cdots\lor a_k \leftarrow b_1,..., b_m, \naf\, b_{m+1},
  \dotsc, \naf\, b_n,
\end{equation}
where each $a_i$, $1 \leq i \leq k$, and each $b_j$, $1 \leq j \leq n$, is a first-order atom.
Given a rule $r$, we call $H(r) = \{a_1,...,a_k\}$ the head of $r$,
$B^+(r) = \{b_1,...,b_m\}$ the positive body of $r$, and $B^-(r) = \{b_{m+1},...,b_n\}$ the negative body of $r$.
A rule is
a \emph{fact} if $n = 0$, a \emph{constraint} if $k = 0$, and
\emph{definite} if $k=1$ and $m = n$. A \emph{definite program} is a
logic program that contains only definite rules.

The grounding $grd(r)$ of a rule $r$ is obtained
as usual;
the grounding of a program $P$ is $grd(P) = \bigcup_{r \in P}grd(r)$. An interpretation $I$ models a ground rule $r$, denoted $I \models r$, if $I \models a_i$ for some $a_i \in H(r)$, whenever $I \models b_i$ for all $b_i \in B^+(r)$ and $I \not\models b_j$ for all $b_j \in B^-(r)$.
An interpretation $I$ models a logic program $P$, denoted $I \models P$, if $I \models r$ for all $r \in grd(P)$; and a definite program $P$ entails a ground atom $a$, denoted $P \models a$, if for every $I$ s.t.\ $I \models P$ it holds that $I \models a$.

\leanparagraph{Meta-Interpretive Learning}
The \emph{Meta-Interpretive Learning} (\emph{MIL}) approach by \citeN{MuggletonLT15} learns definite logic programs from examples by instantiating so-called \emph{meta-rules}.
Here, we 
focus on meta-rules of the form
\begin{equation}
  P(x,y) \leftarrow Q_1(x_1,y_1),...,Q_k(x_k,y_k),R_1(z_1),...,R_n(z_n),
\label{eqn:metarule}
\end{equation}
where $P$, $Q_i$, $1 \leq i \leq k$, and $R_j$, $1 \leq j \leq n$, are higher-order variables, and $x$,$y$,$x_i$,$y_i$, $1 \leq i \leq k$, and $z_j$, $1 \leq j \leq n$, are first-order variables s.t.\ $x$ and $y$ also occur in the body. 
That is, we 
consider meta-rules 
with binary atoms in the head and with binary and/or unary atoms in the body.
Meta-rules with unary head atoms can be simulated by using atoms of
the form $p(x,x)$,
and we allow meta-rules of arbitrary (finite) length,
such that the program class $H^2_{m}$ is covered
(cf.\ \citeANP{CropperM14}, \citeyearNP{CropperM14}).
A \emph{meta-substitution} of a meta-rule $R$ is an instantiation of $R$ where all higher-order variables are substituted by predicate symbols.\footnote{
Even though
we do not consider constants in meta-substitutions, they can easily be simulated
by using e.g.\ a dedicated atom $=_x(x)$ in the body, where $=_x$ is defined in the background knowledge and binds $x$ to a specific constant.}
Examples of concrete meta-rules
with names as used by \citeN{CropperM16} are shown in Figure \ref{tab:metarules}.

\begin{figure}[t]
	\centering
	\begin{tabular}[t]{llll}
		Precon: & $P(x,y) \leftarrow Q(x), R(x,y)$ & Postcon: & $P(x,y) \leftarrow Q(x,y), R(y)$ \\
		Chain: & $P(x,y) \leftarrow Q(x,z), R(z,y)$ & Tailrec: &$P(x,y) \leftarrow Q(x,z), P(z,y)$
	\end{tabular}
	\caption{Examples of Meta-Rules}
	\label{tab:metarules}
\end{figure}

We are now ready to formally introduce the setting of MIL, adapted to our approach.
\begin{definition}
\label{milproblem}
A \emph{Meta-Interpretive Learning} (\emph{MIL-}) \emph{problem} is a
quadruple $\milp = (B,E^+,E^-,\mathcal{R})$,
 where
\begin{itemize}
\item $B$ is a definite program, called  
\emph{background knowledge} (\emph{BK});
\item $E^+$ and $E^-$ are finite sets of binary ground atoms called
\emph{positive} resp.\ \emph{negative examples};
\item $\mathcal{R}$ is a finite set of meta-rules.
\end{itemize}
We say that $B$ is \emph{extensional} if it contains only ground atoms.
A \emph{solution}\/ for
$\milp$ is a
\emph{hypothesis} $\mathcal{H}$
consisting of a set of meta-substitutions of meta-rules in $\mathcal{R}$ s.t.\ $B \cup \mathcal{H} \models e^+$ for each $e^+ \in E^+$ and $B \cup \mathcal{H} \not\models e^-$ for each $e^- \in E^-$.
\end{definition}

In order to obtain solutions that generalize well to new examples,
by Occam's Razor 
simple solutions to MIL-problems are desired; thus Metagol
computes a {\em minimal solution} containing a minimal number of meta-substitutions (i.e.\ rules).

\begin{example}
\label{ex1}
Consider the MIL-problem $\milp$,
with $B = \{m(ann,bob),f(john,bob),$ $m(sue,ann),$ $f(tim,ann)\}$,
$E^+ =\{a(sue,bob),a(tim,bob),$ $a(john,bob)\}$,
$E^- =\{a(bob,tim)\}$, abbreviating $\underline{m}other$, $\underline{f}ather$ and $\underline{a}ncestor$, and meta-rules $\mathcal{R} =
\{P(x,y) \leftarrow Q(x,y);$ $P(x,y) \leftarrow Q(x,z),
R(z,y)\}$.
A minimal solution for $\milp$ is
$\{p1(x,y)$ $\leftarrow$ $f(x,y);$
$p1(x,y)$ $\leftarrow$ $m(x,y);$
$a(x,y)$ $\leftarrow$ $p1(x,y);$
$a(x,y)$ $\leftarrow$ $p1(x,z),$ $a(z,y)\}$, where $p1$
is an invented predicate intuitively representing the concept $parent$.
\end{example}

\citeN{MuggletonLT15} showed that MIL-problems as in Definition~\ref{milproblem} are decidable
if no proper function symbols (i.e., only constants) are used, and $\mathcal{P}$ and $\mathcal{C}$ are finite, but are undecidable in general.
Yet, in practice, complex terms such as lists 
are often used for MIL.  Hence, we assume some
suitable restriction, e.g.\ to consider only a finite set of flat lists, s.t.\ in slight abuse of notation, complex ground terms (e.g., $[a,b,c]$) are
technically
regarded as constants in $\mathcal{C}$.

\leanparagraph{\hex-Programs}
For solving MIL-problems, we exploit the \hex\ formalism \cite{EiterFIKRS16} in our approach.
\hex\emph{-programs} extend disjunctive logic programs by
\emph{external atoms}, which 
can occur in rule bodies.
External atoms are of the form
$\ext{g}{X_1,...,X_k}{Y_1,...,Y_l}$,
where
$X_i \in \mathcal{P} \cup \mathcal{C} \cup \mathcal{X}_1$, $1 \leq i \leq k$, are \emph{input parameters},
and
$Y_j \in \mathcal{C} \cup \mathcal{X}_1$, $1 \leq j \leq l$, are
 \emph{output parameters}.
The semantics of a ground external atom $\ext{g}{p_1,...,p_k}{c_1,...,c_l}$
with $k$ input and $l$ output parameters
wrt.\ an interpretation $I$ is determined by a $1{+}k{+}l$-ary
(\emph{Boolean}) \emph{oracle function} $\extfun{g}$ such that
$I \models \ext{g}{p_1,...,p_k}{c_1,...,c_l}$ iff
$\extsem{g}{I}{p_1,...,p_k}{c_1,...,c_l} = 1$.
In practice, oracle functions are usually realized as plugins
provided to a solver, implemented in $C{++}$ or $Python$ code.

\hex-programs $\Pi$ are interpreted under the \emph{answer set semantics}
\cite{gelf-lifs-91} based on the \emph{FLP-reduct}
by \citeN{flp2011-ai} (a variant of the well-known
\emph{GL-reduct}), 
which for  
an interpretation $I$ is
$f\Pi^{I} = \{ r \in grd(\Pi) \mid I \models b_i \mbox{ for all } b_i
\in B^+(r), 
I \not\models b_j \mbox{ for all } b_j \in
B^-(r) \}$. 
An interpretation $I$ is an \emph{answer set} of a \hex-program $\Pi$
if $I$ is a subset-minimal model of
$f\Pi^I$.

\begin{example}
\label{ex2}
Consider the \hex-program $\Pi = \{l([a,a]);$ $l(y) \leftarrow \ext{remove}{x}{y}, l(x)\}$, and suppose that the oracle function $\extsem{remove}{I}{X}{Y}$
evaluates to $1$ iff $X$ and $Y$ are ground lists and $Y$ can be obtained from $X$ by removing the first list element. The single answer set of $\Pi$ is $\{l([a,a]),l([a]),l([])\}$.
\end{example}

Note that in Example~\ref{ex2}, the output of the external atom
contains constants not occurring in $\Pi$ and are thus
introduced by the external source by so-called
\emph{value invention}. By employing suitable
safety conditions, it can be ensured that only finitely many new
constants 
must be considered. For more details we refer to
\cite{EiterFKR16}.
As \hex\ allows for predicate input to external
atoms, their semantics may depend on the extension
of predicates in an answer set.

\begin{example}
Consider the \hex-program $\Pi = \{p(a) \vee p(b);$ $\leftarrow \ext{contains}{p,b}{}\}$, and suppose that the oracle function $\extsem{contains}{I}{p}{b}$ evaluates to $1$ iff $p(b) \in I$.
Without the constraint, $\Pi$ has the two answer sets $\{p(a)\}$ and $\{p(b)\}$, whereby the constraint eliminates the second one.
\end{example}

\section{\hex-Encoding of Meta-Interpretive Learning}
\label{sec:enc}
In this section, we introduce our main encoding for solving general
MIL-problems, where the \BK\ is stored externally and
interfaced by means of external atoms.  Subsequently, we present a
modification of the encoding which reduces the number of constants
that need to be considered during grounding in case only a certain
type of meta-rules is used.

A major motivation for developing an ASP-based
approach 
to solve MIL-problems 
is that constraints
given by negative examples can be efficiently propagated by an ASP solver,
while Metagol checks them only at the end.
This can be shown by simple synthetic examples;
e.g.\ consider the \BK\ 
of facts
$q^j_i(i)$, $q^{11}_i(i)$ and
$q^j_i(0)$, for $1 \leq i,j\leq 10$. 
For the positive examples
$p(1)$, \ldots, $p(10)$ and the negative example
$p(0)$, Metagol
finds no solution 
within one hour
using meta-rule $P(x) \leftarrow Q(x)$. In contrast,
the problem can be solved by a simple ASP encoding 
 instantly. The reason is that e.g.\ $p(1)$
can only be derived by the rule $p(x) \leftarrow q^{11}_1(x)$
given the negative example $p(0)$,
and Metagol explores a huge number of rule combinations before
this is detected.

While the issue of negative examples can be tackled by
using ordinary ASP, we employ here 
\hex-programs as 
they enable us to outsource the \BK\ from the
encoding. 
This allows us to conveniently specify intensional
BK using, e.g.\ string or list manipulations, which
are usually not available in ASP.  Another advantage of outsourcing
the BK is
that the approach
becomes parametric 
wrt.\  the
formalization of the BK, 
as it is in principle possible to plug in
arbitrary (monotonic) external theories (e.g.\ a description logic ontology).
Beyond this flexibility provided by \hex, external atoms are
essential to limit the \BK\ that is imported as described in the latter part
of this section, and
for realizing our state abstraction technique  
in Section~\ref{sec:abst}.

As we consider meta-rules using unary and binary atoms, we 
introduce 
external atoms for importing the relevant unary and binary atoms
that are entailed by the BK in an encoding.

\begin{definition}
\label{bkatoms}
Given a MIL-problem $\milp$,
we call the external atom $\ext{bkUnary}{deduced}{X,Y}$ \emph{unary BK-atom} and
$\ext{bkBinary}{deduced}{X,Y,Z}$
\emph{binary BK-atom},
where the associated oracle functions fulfill
$\extsem{bkUnary}{I}{deduced}{X,Y}=1$ iff $B \cup \{
p(a,b) \mid deduced(p,a,b) \in I\} \models X(Y)$, respectively
$\extsem{bkBinary}{I}{deduced}{X,Y,Z}=1$ iff $B \cup \{
p(a,b) \mid deduced(p,a,b) \in I\} \models X(Y,Z)$.
\end{definition}

The 
BK-atoms receive as input the extension of the predicate $deduced$,
which represents the set of all atoms that can be deduced from the program that results from the
meta-substitutions of the current hypothesis.
Their output constants represent
unary, resp., binary atoms that are entailed by the \BK\
augmented with the atoms
described by $deduced$.

In theory, MIL can be encoded by applying the
well-known \emph{guess-and-check} methodology,
i.e.\ by generating all combinations of meta-substitutions
from the given meta-rules and available predicate symbols,
deriving all entailed atoms, and checking compatibility with examples using constraints. 
However, this 
results in a huge search space due to the 
many possible
combinations of meta-substitutions, on top of many meta-substitutions
that can be generated by different combinations of predicate
symbols.  At the same time, a large fraction of meta-substitutions is
irrelevant for inducing a hypothesis as the resulting rule bodies can never be
satisfied based on atoms that are deduced using other rules
from the hypothesis and the BK.

For this reason, we interleave guesses on the meta level and
derivations on the object level, i.e.\ deductions using
meta-substitutions already guessed to be part of the hypothesis, and
we model a procedural bias ensuring that meta-substitutions can only
be added if their body is already satisfied by atoms deducible on the
object level.  Note that while 
Metagol's top-down mechanism  
effects that only meta-substitutions necessary for deriving a
goal atom are generated, our approach works bottom-up such that the
procedural bias is inverted.  Guarding the guesses of
meta-substitutions in this way has not been considered by
\citeN{MuggletonLPT14}; this
constitutes the basis for
techniques that restrict the size of the grounding discussed later on.

As in the Metagol implementation of MIL \cite{MuggletonLT15},
given a MIL-problem $\milp = (B,E^+,E^-,\mathcal{R})$,
we associate each meta-rule $R \in \mathcal{R}$ with
a unique identifier $R_{id}$ and a set of \emph{ordering constraints}
$R_{ord} \subseteq \{ord(P,Q) \mid P, Q \in \mathcal{X}_2 \text{ occur in } R\}$; and we assume a predefined total
ordering $\succeq_\mathcal{P}$ over the predicate symbols in $\mathcal{P}$.
The ordering constraints can be utilized to constrain the search space, and are necessary in Metagol in order to ensure termination.
A meta-substitution of a meta-rule $R$ with head predicate $p$ instantiated for the
higher-order variable $P$ satisfies the ordering
constraints $R_{ord}$ in case $p \succeq_\mathcal{P} q$ for every binary body predicate $q$
instantiated for a higher-order variable $Q$
s.t.\ $ord(P,Q) \in R_{ord}$. Here, we apply ordering constraints only to pairs of head and body predicates, but in general this can be extended to arbitrary pairs of predicates in a meta-substitution.
Moreover, we assume that a set $\mathcal{S} \subseteq \mathcal{P}$ of
\emph{Skolem predicates} can be used for predicate invention, where no element in $\mathcal{S}$ occurs in $\mathcal{M}$.

We are now ready to present our main encoding for solving MIL-problems using \hex.

\begin{definition}
\label{hexmilenc}
Given a MIL-problem $\milp = (B,E^+,E^-,\mathcal{R})$ and a finite set of Skolem predicates $\mathcal{S}$,
let $Sig$ be the set that contains each $p \in \mathcal{S}$ and
each predicate symbol $p$ that occurs either in $E^+ \cup E^-$
or in a rule head in $B$.
The \emph{\hex-MIL-encoding} for $\milp$ is the
\hex-program $\Pi(\milp)$
containing
\begin{itemize}
\item[(1)] a fact $sig(p) \leftarrow$ for each $p \in Sig$, and a fact $ord(p,q) \leftarrow$ for all $p,q \in Sig$ s.t.\ $p \succeq_{\mathcal{P}} q$
\item[(2)] the rules~ $unary(x,y) \leftarrow
   \ext{bkUnary}{\mathit{deduced}}{x,y}$ and\\
   \phantom{the rules~} $\mathit{deduced}(x,y,z) \leftarrow \ext{bkBinary}{\mathit{deduced}}{x,y,z}$
\item[(3)] for each meta-rule $R = P(x,y) \leftarrow Q_1(x_1,y_1),...,Q_k(x_k,y_k),R_1(z_1),...,R_n(z_n) \in \mathcal{R}$ \\ and $\{ord(P,Q_{i_1}),...,ord(P,Q_{i_m})\} = \big\{ord(P,Q_i) \in R_{ord} \mid i \in \{1,...,k\}\big\}$,
\begin{itemize}
\item[(a)] a rule
\begin{align*}
m&eta(R_{id},x_P,x_{Q_1},...,x_{Q_k},x_{R_1},...,x_{R_n}) \vee n\_meta(R_{id},x_P,x_{Q_1},...,x_{Q_k},x_{R_1},...,x_{R_n}) \leftarrow\\
&sig(x_P), sig(x_{Q_1}), ..., sig(x_{Q_k}), sig(x_{R_q}), ..., sig(x_{R_n}), ord(x_{P},x_{Q_{i_1}}),...,ord(x_{P},x_{Q_{i_m}}),\\
&\mathit{deduced}(x_{Q_1},x_1,y_1), ..., \mathit{deduced}(x_{Q_k},x_k,y_k), unary(x_{R_1},z_1), ..., unary(x_{R_n},z_n)
\end{align*}
\item[(b)] and a rule
\begin{align*}
d&educed(x_P,x,y) \leftarrow meta(R_{id},x_P,x_{Q_1},...,x_{Q_k},x_{R_1},...,x_{R_n}),\\
&\mathit{deduced}(x_{Q_1},x_1,y_1), ..., \mathit{deduced}(x_{Q_k},x_k,y_k), unary(x_{R_1},z_1), ..., unary(x_{R_n},z_n)
\end{align*}
\end{itemize}
\item[(4)] a constraint~~ $\leftarrow \naf \mathit{deduced}(p,a,b)$ for each $p(a,b) \in E^+$, and\\ a constraint~~ $\leftarrow \mathit{deduced}(p,a,b)$ for each $p(a,b) \in E^-$
\end{itemize}
\end{definition}

In the encoding, the predicate $meta$ contains meta-substitutions
added to an induced hypothesis, and $deduced$ captures
all atoms that can be deduced from a guessed hypothesis together with
the BK. As we consider examples to be binary atoms and only binary
atoms can be derived from meta-substitutions, those binary atoms entailed by the \BK\ are directly derived to be in the extension of $deduced$, while unary atoms can only be derived from the \BK\ s.t.\ they do not need to be added to the extension of $deduced$ and are imported via the predicate $unary$ in item (2).

Item (3) constitutes the core of the encoding, which contains the meta-level guessing part (a) and the object-level deduction part (b). A meta-substitution can be guessed to be part of the hypothesis only if first-order instantiations of its body atoms can already be deduced, i.e.\ only if it is potentially useful for deriving a positive example. At this, predicate names must be from the signature $Sig$ and the ordering constraints must be satisfied as stated by the facts in item (1).
Finally, item (4) adds the constraints imposed by the positive and negative examples.

For a given MIL-problem, solutions constituted by induced logic programs can directly be obtained from the answer sets of the respective \hex-MIL-encoding.
The induced logic program represented by the $meta$-atoms in an interpretation is extracted as follows:

\begin{definition}
For a set of meta-rules $\mathcal{R}$, the \emph{logic
  program induced by} a
given interpretation $I$ consists
of all rules obtained
from an atom of the form
$meta(R_{id},x_P,x_{Q_1},...,x_{Q_k},x_{R_1},...,x_{R_n})$ in $I$ such
that the meta-rule $R = P(x,y) \leftarrow
Q_1(x_1,y_1),...,Q_k(x_k,y_k),R_1(z_1),...,R_n(z_n)$ is in $\mathcal{R}$, by substituting $P$ by $x_P$, $Q_i$ by $x_{Q_i}$ for $1 \leq i \leq k$, and $R_j$ by $x_{R_j}$ for $1 \leq j \leq n$.
\end{definition}

In the following, we assume that $\mathcal{R}$
is given by the respective MIL-problem at hand. 

Every answer set of a \hex-MIL-encoding encodes a solution for the
respective MIL-problem, and all solutions $\mathcal{H}$ that only contain \emph{productive} rules, i.e.\ rules such that all atoms in the body of some ground instance is entailed by $B \cup \mathcal{H}$, can be generated in this way.

\begin{theorem}
Given a MIL-problem $\milp$, 
(i) if $S$ is an answer set of $\Pi(\milp)$, the logic program $\mathcal{H}$ induced by $S$ is a solution for $\milp$; and (ii)
if $\mathcal{H}$ is a solution for $\milp$ s.t.\ 
all rules in $\mathcal{H}$ satisfy $R_{ord}$ and are productive,
then there is an answer set $S$ of $\Pi(\milp)$ s.t.\ $\mathcal{H}$ is the logic program induced by $S$.
\end{theorem}

Although the general \hex-MIL-encoding 
in Definition \ref{hexmilenc}
works well when only a small number of constants is introduced by
the BK-atoms, the grounding
can quickly become prohibitively large when many constants are generated
(e.g.\ due to list operations). This results from the fact that constants
produced by item (2) in Definition \ref{hexmilenc} are also relevant
for instantiating the rules defined in items (3a) and (3b),
which contain many variables, causing a combinatorial explosion.

\begin{example}
\label{ex4}
Consider a MIL-problem $\milp$,
containing \BK\ $B = \{remove([X|R],R)\leftarrow\}$,
and the positive examples $E^+ =$ $\{remove2([a,a,a],[a]), remove2([b,b],[])\}$.
Here, the definition of the \BK\ should be read as 
an abbreviation for a set of facts, e.g.\ containing $remove([a,a],[a])$, $remove([a],[])$, etc.,
exploiting the list notation of Prolog.
Accordingly, the predicate $remove$ drops the first element from a list,
and a corresponding hypothesis intuitively needs to remove the first two elements
from the list in the first argument of an example to yield the second one.

Now, assume that $\mathcal{C}$ contains lists with letters from the set
$\{a,b,c\}$ up to some length $n$.
Then, the \BK\ contains, e.g., $remove([c,c],[c])$,
$remove([c,c,c],[c,c])$, etc., up to length $n$, which are imported
via the BK-atoms.
However, lists containing the letter $c$ are irrelevant wrt.\
$\milp$ because they cannot be obtained from lists
appearing in the examples using 
the operations
in the BK.
\end{example}

Next, we introduce a class of meta-rules that allows us to restrict
the number of constants imported from the BK,
based on the observation from the previous example.

\begin{definition}
A \emph{forward-chained} meta-rule is of the form

\smallskip

\centerline{$ P(z_0,z_k) \leftarrow Q_1(z_0,z_1),\dotsc,Q_i(z_{i-1},z_i),\dotsc,Q_k(z_{k-1},z_k),R_1(x_1),\dotsc,R_l(x_l),$}

\smallskip

\noindent where $1 \leq i \leq k$, $0 \leq l$, and $x_j \in \{z_0,\dotsc,z_{k}\}$ for all $1 \leq j \leq l$.
A MIL-problem $\milp$ is \emph{forward-chained} if $\mathcal{R}$ only contains forward-chained meta-rules.
\end{definition}

Intuitively, all first-order variables in the body of a forward-chained
meta-rule are part of a chain between
the first and second argument of the head atom.
Viewing binary predicates in the BK
as mappings from their first to their second argument,
only atoms from an extensional \BK\ are relevant that occur in a chain
between the first and the second argument of examples. Hence, atoms from the \BK\ only need to be imported
when their first
argument occurs in the examples
or in a deduction wrt.\ \BK\ that has already been imported.
However, when the derivable \BK\ depends on guessed
meta-substitutions, additional atoms might be relevant, and thus, we only consider extensional \BK\ in the following.

For restricting the import of BK, we introduce a modification of 
the external atoms from Definition \ref{bkatoms},
where the output is guarded by an input constant.

\begin{definition}
Given a forward-chained MIL-problem $\milp$ where $B$ is extensional,
we call the external atoms $\ext{fcUnary}{Y}{X}$ and
$\ext{fcBinary}{Y}{X,Z}$
\emph{unary} and \emph{binary forward-chained BK-atom}, resp., 
where $\extsem{fcUnary}{I}{Y}{X}=1$ iff $X(Y) \in
B$, resp., $\extsem{fcBinary}{I}{Y}{X,Z} = 1$ iff $X(Y,Z) \in B$.
\end{definition}

As we assume the \BK\ to be extensional, the input parameter $deduced$ is not needed for forward-chained BK-atoms.
Based on the previous definition, we can modify our \hex-MIL-encoding such that only
relevant atoms from the \BK\ are imported,
where forward-chained BK-atoms receive as input all constants that already occur in a deduction or the examples.

\begin{definition}
\label{fcenc}
Given a forward-chained MIL-problem $\milp$ where $B$ is extensional,
the \emph{forward-chained \hex-MIL-encoding} for
$\milp$ is the \hex-program
$\Pi_f(\milp)$
containing items (1), (3) and (4) from Definition~\ref{hexmilenc},
and the rules
\begin{parcolumns}[]{2}
\colchunk{
\begin{itemize}
\item[(f1)] $unary(x,y) \leftarrow \ext{fcUnary}{y}{x}, s(y)$
\item[(f2)] $deduced(x,y,z) \leftarrow \ext{fcBinary}{y}{x,z}, s(y)$
\end{itemize}
}
\colchunk{
\begin{itemize}
\item[(f3)] $s(a) \leftarrow$ for each $p(a,b) \in E^+ \cup E^-$
\item[(f4)] $s(y) \leftarrow deduced(\_,\_,y)$
\end{itemize}
}
\end{parcolumns}
\end{definition}

The main difference between $\Pi_f(\milp)$ and $\Pi(\milp)$
is that
the import of \BK\ is guarded by 
the predicate $s$ in items 
(f1) and (f2), 
whose extension contains all constants appearing as first argument of an example,
due to item (f3), and all constants that 
appear in deductions based on the already imported \BK, due to item (f4).

Every answer set of the forward-chained \hex-MIL-encoding still
corresponds to a solution of the respective MIL-problem, but not all
solutions may be obtained.
Nonetheless, it is ensured that a minimal solution (i.e., with 
fewest meta-substitutions) is encoded by some answer set:

\begin{theorem}
Let $\milp$ be a forward-chained MIL-problem with extensional
$B$.  
Then, (i) for every answer set  $S$ of $\Pi_f(\milp)$, the logic program
induced by $S$ is a solution for $\milp$;
and (ii) there is an answer set $S'$ of $\Pi_f(\milp)$ s.t.\ the logic
program 
induced by $S'$ is a minimal solution for
$\milp$ if one exists.
\end{theorem}

Since, in practice, we employ iterative deepening search
for computing a minimal solution, any minimal solution encoded by an
answer set of $\Pi_{f}(\milp)$ is guaranteed to be found.
Thus, 
 we can obtain minimal solutions
 while
grounding issues are mitigated by steering
the import of BK. 
An additional search space reduction results from the pruning of 
the grounding.

\section{State Abstraction}
\label{sec:abst}
Based on the  
observation that operations represented by
binary \BK\ predicates can be applied sequentially when only
forward-chained meta-rules are used, we introduce in this
section 
a further technique  
that eliminates object-level constants from the encoding entirely.
While the 
$\Pi_f(\mathcal{M})$-encoding
 focuses
the import of
constants to those  
obtainable from constants that already occur in deductions,
the number of relevant constants can still be large 
if many binary
\BK\ atoms 
share the first argument; and all of them 
must be considered during grounding.
However, only one \BK\ atom is needed for each element in
a chain that derives a positive example
$p(x,y)$ by connecting 
$x$ and $y$.
In fact, the 
$\Pi_f(\mathcal{M})$-encoding solves two problems at the same time:
(1) finding 
sequences of binary \BK\ predicates that
derive positive examples;
and (2)
inducing a (minimal) program that 
calls the predicates in the
respective sequences, and prevents the derivation of
negative examples.

\begin{example}
\label{ex6}
Consider the MIL-problem $\milp$ where $B$ contains
the extension of
$remove$ from Example~\ref{ex4}, and extensional \BK\ represented by
$switch([X,Y|R],[Y,X|R])\leftarrow$ and $firstA([a|R])\leftarrow$. Furthermore, let $E^+ =$
$\{p([c,a,b,a,b],[c])\}$, $E^- =$ $\{p([c,b,a,b,b],[c])\}$, and
$\mathcal{R} = \{P(x,y) \leftarrow Q(x,z), R(z,y);$ $P(x,y) \leftarrow Q(x,y),
R(y);$ $P(x,y) \leftarrow$ $Q(x,y)\}$.  Intuitively, a
solution program needs to memorize $c$ and delete the rest; this
requires to repeatedly switch the first two elements and remove the
first element. For success, the input list must have $a$ at
position~2. 
This is captured by the hypothesis $\mathcal{H} = \{ p(x,y) \leftarrow$
$p1(x,z),$ $p(z,y);$ $p(x,y)$
$\leftarrow$ $\mathit{remove}(x,y);$ $p1(x,y)$ $\leftarrow$
$\mathit{switch}(x,y),$ $firstA(y);$ $p1(x,y)$ $\leftarrow$ $\mathit{remove}(x,z),$
$\mathit{switch}(z,y)\}$, 
where $p1$ is an invented
predicate; this is in fact a minimal solution for $\milp$. 
In addition, any program which enables derivations that alternate
between calling  
$\mathit{switch}$ and $\mathit{remove}$ and prevents to
derive the negative example
using $firstA$ as  
a guard would 
be a solution.
Notably, the search
space of Metagol also contains hypotheses 
that 
have no
alternation between $\mathit{switch}$ and $\mathit{remove}$ and thus
cannot be solutions.
\end{example}

The previous example illustrates that the derivability of positive
examples depends on the sequences 
by which binary \BK\ predicates are
called in the induced program. 
Here, finding a correct sequence for
a given example can be viewed as a \emph{planning problem}, where
object-level constants represent \emph{states}, binary \BK\ predicates
are 
viewed as \emph{actions}, and unary \BK\ predicates
constitute \emph{fluents}.  The \emph{state abstraction} technique
described in the 
sequel exploits the insight that
the tasks of (1) solving the planning problem and (2) finding a matching
hypothesis can be separated, where the \hex-program encodes 
task~(2), and computations involving states are performed externally.
The advantage of task separation and state abstraction increases with
the number of
actions that 
are applicable in a state, as usually more actions not occurring
in a derivation of a positive example can be ignored; 
this reduces the search space 
and the size of the grounding.

We represent possible plans 
to derive positive examples by
sequences of
binary 
\BK\ atoms.
At this,
cyclic sequences (or plans) have to be excluded
by requiring that 
constants (states) occur only once because otherwise, we may obtain infinitely many sequences for a positive example:

\begin{definition}
\label{def:sequences}
Given a forward-chained MIL-problem $\milp$ where $B$ is extensional,
the function $Seq$ maps each positive example $p(c_1,c_{k}) \in E^+$ to the set $Seq(p(c_1,c_{k}))$ containing
all sequences $p_1(c_1,c_2),\dotsc,$ $p_{k-1}(c_{k-1},c_{k})$, where $p_i(c_i,c_{i+1}) \in B$ for all $1 \leq i < k$, and $c_i \not= c_{j}$ if $i \not= j$.
\end{definition}

\begin{example}[cont'd]
\label{ex7}
Reconsider $\milp$ from Example~\ref{ex6}.
Then $Seq(p([c,a,b,a,b],[c])) = \{ seq \}$, ($s$ = $\mathit{switch}$, $r$ = $\mathit{remove}$),
\begin{align*}
seq ~=~ &s([c,a,b,a,b],[a,c,b,a,b]), r([a,c,b,a,b],[c,b,a,b]),
s([c,b,a,b],[b,c,a,b]),\\
&r([b,c,a,b],[c,a,b]), s([c,a,b],[a,c,b]), r([a,c,b],[c,b]),s([c,b],[b,c]), r([b,c],[c]).
\end{align*}
\end{example}

In order to make information about action sequences that derive positive
examples and fluents that hold 
in states available to the \hex-encoding, we next introduce two external atoms
that
import 
such information. States
are simply represented by integers in the output 
as their structure
is
irrelevant for combining 
sequences into a hypothesis that generalizes the plans.

\begin{definition}
\label{abstractimport}
For a forward-chained MIL-problem $\milp$ where $B$ is extensional, let
$e^+_{id}$ and $seq_{id}$ be unique identifiers, resp., 
for each $e^+ \in E^+$ and $seq \in \bigcup_{e^+ \in E^+} Seq(e^+)$.
The external atoms $\ext{saUnary}{}{X,Y}$ and
$\ext{saBinary}{}{X,Y,Z}$ are called
\emph{unary} and \emph{binary state abstraction}
(\emph{sa-})\emph{atoms}, resp., 
where
\begin{itemize}
\item  
$\extsem{saUnary}{I}{X}{Y}=1$ iff $X = r$, $Y = (e^+_{id},seq_{id},i)$, and $r(c_i) \in B$; resp.
\item
$\extsem{saBinary}{I}{X}{Y,Z}=1$ iff $X = p_i$, $Y = (e^+_{id},seq_{id},i)$, and $Z = (e^+_{id},seq_{id},i+1)$,
\end{itemize}
with $e^+ \in E^+$, $seq = p_1(c_1,c_2),\dotsc,p_{k-1}(c_{k-1},c_{k})
\in Seq(e^+)$, and $i\in \{1,\ldots, k-1\}$.
\end{definition}

For instance, for $\milp$ from Example~\ref{ex6},
$\ext{saBinary}{}{switch,(e^+_{id},seq_{id},1),(e^+_{id},seq_{id},2)}$ is true,
where $e^+_{id}$ is the identifier of the positive example, $seq_{id}$ is the identifier of
the sequence shown in Example~\ref{ex7}, and the integers $1$ and $2$ represent the states
$[c,a,b,a,b]$ and $[a,c,b,a,b]$, respectively, where the second state can be reached from the
first state by applying the action $switch$.

In our
encoding with state abstractions                  
we also need information about the start and end states of sequences associated with positive
examples, as a
hypothesis needs to encode a plan for each positive example.  This
information is accessed 
via an external atom as well.
\begin{definition}
For a forward-chained MIL-problem $\milp$,
the external atom $\ext{checkPos}{}{X_1,X_2,Y,Z}$
fulfills that
$\extsem{checkPos}{I}{X_1,X_2}{Y,Z}=1$ iff
$X_1 = e^+_{id}$, $X_2 = p$, $Y = (e^+_{id},seq_{id},1)$, and $Z =
(e^+_{id},$ $seq_{id},k\,{+}\,1)$
for some $p(a,b) = e^+ \in E^+$ and
$seq = p_1(a,c_2),\dotsc,$ $p_k(c_k,b)$ $\in Seq(e^+)$.
\end{definition}

Finally, it can only be determined wrt.\ the \BK\ whether
a candidate hypothesis 
derives a negative example, s.t.\
the corresponding check cannot be performed in an encoding without importing
relevant atoms from the BK.
As our goal is to abstract from explicit states in the \BK,
we also need to outsource the check for non-derivability of negative examples
by an external constraint.

\begin{definition}
Given a MIL-problem $\milp$,
the oracle function $f_{\&failNeg}(I,meta)$ associated with
the external atom $\ext{failNeg}{meta}{}$ evaluates to $1$ iff
$B \cup \mathcal{H} \models e^-$
for some $e^- \in E^-$, where $\mathcal{H}$ is the logic
program induced by
$\{meta(R_{id},x_P,x_{Q_1},...,x_{Q_k},x_{R_1},...,x_{R_n}) \in I\}$.
\end{definition}

In the implementation, the external atom $\ext{failNeg}{meta}{}$ receives information
about meta-substitutions already guessed by the solver
to be in the respective hypothesis. It can 
be evaluated to true as soon
as a negative example is derivable wrt.\ its input,
as definite logic programs are monotonic;
as this may violate a constraint,  backtracking in a solver can be triggered.
\begin{example}
\label{ex7-A}
Consider MIL-problem
$\milp$
with $B = \{q(a,b),q(a,c),r(a,b)\}$,
$E^+ = \{p(a,b)\}$,
$E^- = \{p(a,c)\}$, and
$\mathcal{R} = \{R = P(x,y) \leftarrow Q(x,y)\}$.
For $I = \{meta(R_{id},p,q)\}$, we obtain 
that 
$f_{\&failNeg}(I,meta) = 1$
as
the negative example can be derived from $B \cup \{p(x,y) \leftarrow
q(x,y)\}$;
a solver can exploit the information that $p(x,y) \leftarrow q(x,y)$
cannot
belong to any solution. 
\end{example}

Utilizing the external atoms introduced in this section, we
define an encoding 
which
separates the planning from the generalization problem
and contains no object-level constants.
\begin{definition}
\label{saencoding}
Given a forward-chained MIL-problem $\milp$ where $B$ is extensional,
its \emph{state abstraction} (\emph{sa-}) \emph{\hex-MIL-encoding}
is the \hex-program $\Pi_{sa}(\milp)$
that contains all rules in items (1) and (3) of
Definition~\ref{hexmilenc}, where $Sig$ additionally contains
$e^+_{id}$ for each $e^+ \in E^+$,
and the rules
\begin{parcolumns}[]{2}
\colchunk{
\begin{itemize}
\item[(s1)] $\mathit{unary}(x,y) \leftarrow \ext{saUnary}{}{x,y}$
\item[(s2)] $\mathit{deduced}(x,y,z) \leftarrow \ext{saBinary}{}{x,y,z}$
\item[(s3)] $\leftarrow \naf\  \mathit{pos1}(e^+_{id})$, for each $e^+ \in E^+$
\item[(s4)] $\mathit{pos}(x_{id},x,y,z) \vee \mathit{n\_pos}(x_{id},x,y,z) \leftarrow \ext{checkPos}{}{x_{id},x,y,z}$
\end{itemize}
}
\colchunk{
\begin{itemize}
\item[(s5)] $\mathit{pos1}(x_{id}) \leftarrow \mathit{pos}(x_{id},\_,\_,\_)$
\item[(s6)] $\leftarrow \naf\ \mathit{deduced}(x,y,z), \mathit{pos}(\_,x,y,z)$
\item[(s7)] $\leftarrow \ext{failNeg}{meta}{}$
\end{itemize}
}
\end{parcolumns}
\end{definition}

Items (s1) and (s2) in $\Pi_{sa}(\milp)$ import the
fluents for all relevant states and  
state transitions wrt.\ sequences that derive positive examples,
where states are abstracted.
The external atom $\ext{checkPos}{}{X_1,X_2,Y,Z}$ in item (s4) imports all tuples
representing the start and end state of each sequence for each positive example.
The disjunctive head of (s4) enables each tuple representing a sequence to
be guessed to be in the extension of the predicate $pos$, which represents
all sequences that are modeled by the induced program.
While a minimal hypothesis is guaranteed when 
the guess is over all possible sequences,
in practice, we can preselect sequences
returned by the atom $\ext{checkPos}{}{X_1,X_2,Y,Z}$.
Moreover, the guess can be omitted if the planning problem is deterministic,
i.e.\ if for each positive example
there is exactly one sequence of binary atoms from the \BK\ that derives its second argument from its first argument.
Items (s3) and (s5) 
ensure that at least one sequence for
each positive example is selected s.t.\ the corresponding
end state can be derived from the start state by the induced program.
Finally, 
(s6) and (s7) state the constraints
regarding positive resp.\ negative examples. 

As can be shown, $\Pi_{sa}(\milp)$ 
only yields correct solutions, and a minimal one if all sequences that derive positive examples are acyclic. More formally:

\begin{theorem}
Let $\milp$ be a forward-chained MIL-problem with extensional \BK\
$B$. 
Then, (i) for every answer set  $S$ of $\Pi_{sa}(\milp)$, the logic program
induced by $S$ is a solution for $\milp$;
and (ii) there is an answer set $S'$ of $\Pi_{sa}(\milp)$ s.t.\ the logic
program
induced by $S'$ is a minimal solution for
$\milp$ if one exists and every sequence of binary \BK\ atoms that derives a positive example in $E^+$ is acyclic.
\end{theorem}

Hence, we have an alternative  
to find solutions for
forward-chained MIL-problems where  
planning and generalization are separated in a way such that the \BK\ can be
outsourced completely.

\section{Empirical Evaluation}
\label{sec:eval}
In this section, we evaluate our approach by comparing it
to Metagol in terms of efficiency.

\leanparagraph{Experimental Setup}
For experimentation,
we utilized an iterative deepening strategy which incrementally
increases a limit for the maximal number of guessed meta-substitutions imposed
via a constraint
to obtain minimal solutions.
In addition, we incrementally increased the number of invented predicates wrt.\
each limit, which
proved to be beneficial for performance.

\begin{figure}[t]
\includegraphics[width=4.4cm]{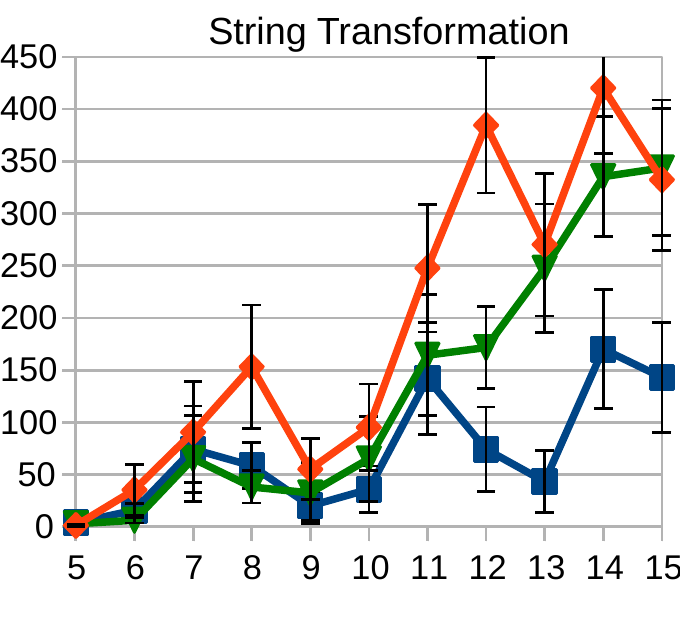}
\includegraphics[width=4.4cm]{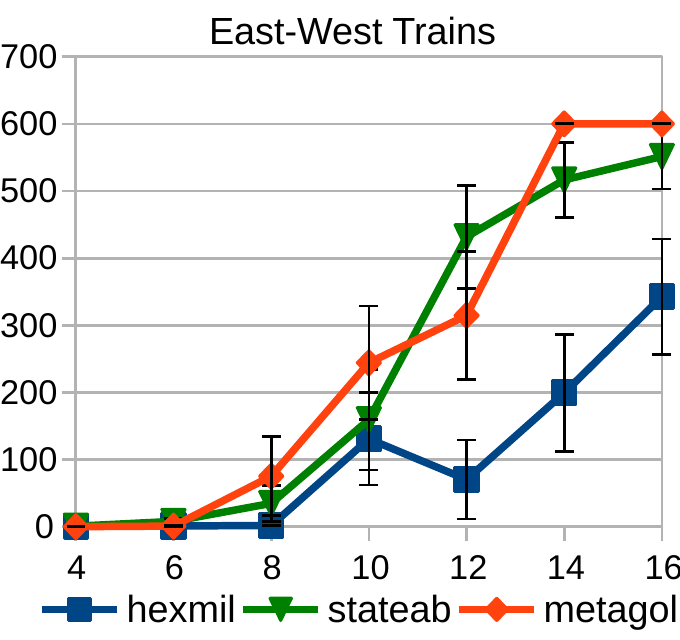}
\includegraphics[width=4.4cm]{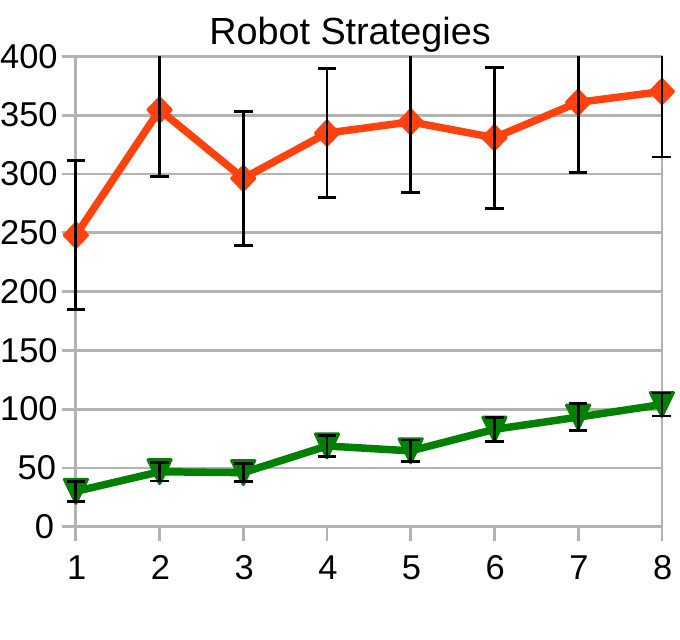}
\caption{Benchmark results for \emph{String Transformation}, \emph{East-West Trains} and \emph{Robot Strategies} (left to right). Average overall running times in seconds are shown on the y-axis, and instance size is shown on the x-axis.}
\label{results}
\end{figure}

\begin{figure}[t]
\includegraphics[width=4.4cm]{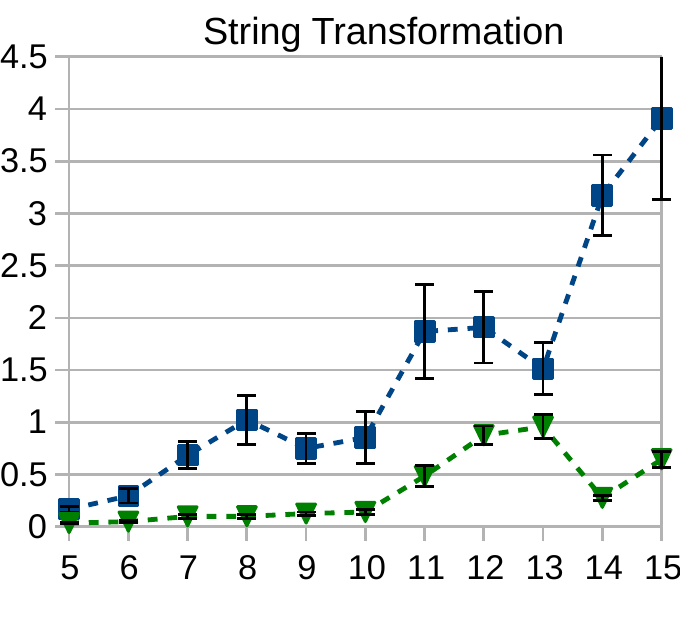}
\includegraphics[width=4.4cm]{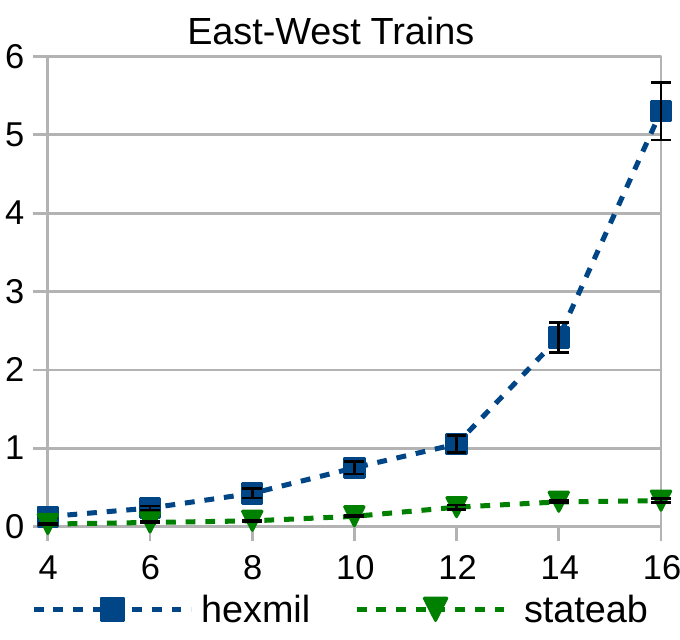}
\includegraphics[width=4.4cm]{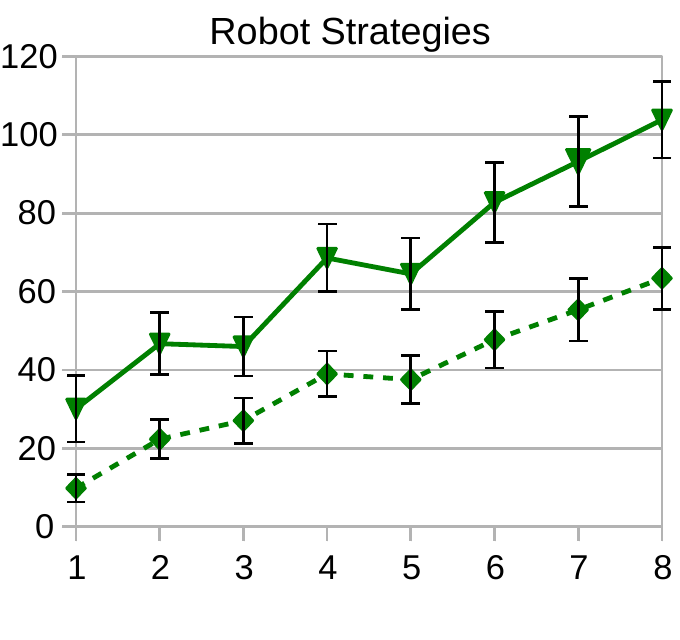}
\caption{Grounding times. Dashed lines indicate the average running times required for grounding in seconds, the solid line in the rightmost diagram shows the overall running times of benchmark B3 for comparison. Overall running times are not shown for benchmarks B1 and B2 as they are much larger than the grounding times. Grounding in condition $hexmil$ was infeasible for benchmark B3.}
\label{results2}
\end{figure}

We computed answer sets of our encodings
with \emph{hexlite}\footnote{https://github.com/hexhex/hexlite/} 
0.3.20,
which is based on \textsc{clingo} 5.1.0.
For comparison, we used \emph{SWI-Prolog} 7.2.3 to run Metagol 2.2.0 \cite{metagol}.
Experiments were run on a Linux machine with
2.5\,GHz dual-core Intel Core i5 processor and 8\,GB RAM;
the timeout was 600 seconds per instance.
The results wrt.\ the average running times in seconds are 
shown in Figure~\ref{results},
where error bars indicate
the \emph{standard error} (= sdvn $s / \sqrt{n}$
for $n$ instances)
per instance size.
In addition, the average running times required for the grounding step are shown
in Figure~\ref{results2}.
We compared the encodings $\Pi_f(\milp)$ and $\Pi_{sa}(\milp)$ (conditions \emph{hexmil} and
\emph{stateab} in Figure~\ref{results}, resp.) to Metagol for the first two benchmarks,
and only used $\Pi_{sa}(\milp)$ for the third benchmark as discussed below.

For each MIL-problem
in this section, we used the meta-rules shown in
Figure~\ref{tab:metarules}, and we implemented it
in Metagol and used our \hex-MIL-encodings.
External atoms are realized as Python-plugins in our implementation.
For operations defined by the \BK, we utilized custom list manipulations.
The external atoms $\ext{checkPos}{}{X_1,X_2,Y,Z}$ and $\ext{failNeg}{meta}{}$ in $\Pi_{sa}(\milp)$ employ breadth-first search for computing all sequences wrt.\ positive examples and for checking the derivability of negative examples, respectively.
In the further development of our implementation, our goal is to employ more sophisticated planning algorithms for computing the sequences, and to interface a Prolog-interpreter for processing the \BK\ and for checking negative examples.

The encodings for the benchmark problems and
all instances used in the experiments are available at http://www.kr.tuwien.ac.at/research/projects/inthex/hexmil/.

\leanparagraph{String Transformation (B1)}
Our first benchmark is based on
Example~\ref{ex6},
and akin 
to inducing
\emph{regular grammars} as considered by \citeN{MuggletonLPT14}.
Learning grammars is a suitable use case for MIL as it enables
recursive string processing
and predicate invention
to represent substrings.
In contrast to \citeN{MuggletonLPT14}, we also
allow switching the first two letters in a string in addition to removing elements, which increases
the search space and makes conflict propagation and state abstraction more relevant.
For the instances used by \citeN{MuggletonLPT14}, Metagol performs much
better due to limited branching in the search space.
We used positive and negative examples of the form $p([c|X],[c])$, where $X$ is a random sequence of letters $a$ and $b$.
The predicates in the \BK\ are
$\mathit{remove}$, $\mathit{switch}$, $\mathit{firstA}$, $\mathit{firstB}$ and $\mathit{firstC}$ (cf.\ Example \ref{ex6}).
We used problems containing one positive
and one negative example of the same length, and
tested lengths $n \in \{1,...,15\}$. 
We report average runtimes of
20 randomly generated instances per $n$.

\leanparagraph{East-West Trains (B2)}
The \emph{East-West train challenge}
by \citeN{LarsonM77}
is a popular ILP-benchmark.
The task is to learn a theory that classifies trains based
on features
(e.g.\ shapes of cars and types of loads) to be either east- or westbound.
In our benchmark, eastbound trains are
positive and westbound trains negative examples, where 
trains are represented by lists. The \BK\ defines the operation $\mathit{removeCar}$
which removes the first car from a train; and we declare
50 different unary predicates, e.g.\ $\mathit{shape\_rectangle}$ or $\mathit{load\_3\_triangles}$,
for checking properties of the remaining part of a train.
We used a data set
of 10 eastbound and 10 westbound trains  
proposed by
\citeN{michie1994international} 
that was also
considered 
by \citeN{MuggletonLT15}.
We generated instances of size $n \in \{4,6,8,10,12,14,16\}$
by randomly selecting $n$ from the 20 trains, s.t.\ $n/2$ were eastbound, and
averaged the running times of 10 instances for each problem size.

\leanparagraph{Robot Waiter Strategies (B3)}
For our final experiment, we
used a problem 
by \citeN{CropperM16} that consists in learning robot strategies:
customers
sit at a table in a row, and a waiter robot
serves each customer her desired drink, which is either tea or coffee.
Initially, the robot is
at the left
end of the table and each customer has an empty cup.
In the goal state, 
each cup
contains the desired drink and the robot
is
at the right end of the table.
States are represented using lists,
and positive examples map  
an initial state to a goal state considering
different numbers of customers and preferences for drinks.
The actions 
are defined by binary \BK\ predicates 
$\mathit{move\_right}$, 
$\mathit{pour\_coffee}$ and $\mathit{pour\_tea}$,
and 
the fluents  
by unary \BK\ predicates  
$\mathit{wants\_coffee}$, $\mathit{wants\_tea}$ and
$\mathit{at\_end}$.\footnote{In contrast to \citeN{CropperM16},
we omitted the action $\mathit{turn\_cup\_over}$, as otherwise we obtained timeouts for the majority of
instances and all conditions, as it is also the case for Metagol in \cite{CropperM16}.}
A solution constitutes a planning strategy by
generalizing a plan for each positive example.

For this benchmark, solutions
are constrained to be \emph{functional}, i.e.\
to map an initial state only
to the unique respective goal state and not to any non-goal state.
Accordingly, negative examples are 
implicitly given by
all binary atoms that map 
an initial state to 
a non-goal state.
In Metagol, solutions can be restricted to functional theories by means
of a property declaration, and we also integrated a corresponding check in the implementation for the
external atom 
$\ext{failNeg}{meta}{}$.

We
generated random instances similar to \citeN{CropperM16},
where each positive example has a random number of 
$i \in [1,10]$
customers
 with random drink preferences, and the instance
size is measured in terms of the number of positive examples ranging from 1 to 8. For each instance size we averaged the running times of 20 problem instances.

\leanparagraph{Findings}
Regarding (B1) and (B2), we found that instances can be solved
significantly faster by employing 
$\Pi_f(\milp)$ than by Metagol due to conflict propagation in ASP.
The encoding $\Pi_{sa}(\milp)$ performed similar to Metagol since only two binary predicates, resp.\ one, are defined by the \BK\ s.t.\ solving the
planning problem externally does not yield a significant advantage, and the advantage of efficient conflict propagation in ASP is outweighed
by the overhead that goes along with outsourcing constraints for negative examples
in $\Pi_{sa}(\milp)$.
$\Pi_{sa}(\milp)$ performs slightly better in
(B1), where two actions are available instead of only one in (B2).

For (B3), we did not obtain results by using
$\Pi_f(\milp)$ for many instances as the grounding was too large due to
 the imported \BK. For instance size 5, the import
from the \BK\ already consumed around 100 MB of memory due to the high
number of states, and the grounding
of the encoding exceeded the available memory.
However, the grounding problem could effectively be avoided by using state
abstractions with $\Pi_{sa}(\milp)$, which yielded a significant speed-up
compared to Metagol. This is due to the fact that by using $\Pi_{sa}(\milp)$,
the planning problem is split from the generalization problem such
that only one precomputed plan
per positive example is considered,
which greatly reduced the search space.
Overall, the performance could be improved by one of our encodings wrt.\ Metagol in all benchmarks,
whereby state abstraction was crucial when many different actions are defined by the \BK, but may decrease efficiency otherwise.

With respect to 
the grounding step,
we found that grounding $\Pi_{f}(\milp)$ required significantly
more resources in terms of running time as well as the size of the grounding
than grounding $\Pi_{sa}(\milp)$, both in the case of (B1) and (B2).
The reason is that
only states need to be considered which occur in a sequence
of binary \BK\ atoms that derives a positive example for grounding the encoding $\Pi_{sa}(\milp)$,
while $\Pi_{f}(\milp)$ also imports all constants that are potentially relevant for deriving some negative example.
However, the advantage of $\Pi_{sa}(\milp)$ wrt.\ the grounding step is 
canceled out for (B1) and (B2) due to the overhead that goes along with outsourcing the check for negative examples and the small advantage in terms of search space pruning. Moreover, the grounding time required for (B1) and (B2) in general is negligible compared to the solving time.
In contrast, we observed that the running time required for grounding $\Pi_{sa}(\milp)$ in the case of (B3) makes up more than
half of the overall running time. This is explained by the fact that the  external atoms $\&checkPos$, $\&saUnary$ and $\&saBinary$ need to be evaluated during grounding due to value invention, which accounts for the major fraction of the overall grounding time.

We also tested the effect of fixing the number of invented predicates,
and obtained timeouts for many instances which could be solved otherwise. The reason is that the availability of additional predicate symbols blows up the search space at the last iteration of the iterative deepening search.
Consequently, the advantage of finding solutions with fewer invented predicates faster compensates for the time invested in restarting the solver many times during iterative deepening.

\section{Discussion}
\label{sec:concl}
In this section, we first discuss
the practical implications of constraining the shapes of meta-rules
that can be employed for learning, and the limitations of our state abstraction technique.
Additionally, we discuss some possible mitigations wrt.\ these limitations, which
are the subject of future work.
Previous work related to our approach is discussed at the end of this section.

\leanparagraph{Meta-Rules}
In this paper, we focused on meta-rules according to Equation~\ref{eqn:metarule},
and restricted the form of meta-rules and of the \BK\ for the encodings $\Pi_{f}(\mathcal{M})$ and $\Pi_{sa}(\mathcal{M})$.
At this, the fragment of \emph{dyadic Datalog}, i.e.\ the class of Datalog programs with predicates of arity at most two, is extremely important in practice as it is suitable whenever
input data is given in form of a graph.
Moreover, the seminal paper on MIL mainly focused on the program class $H^2_2$ and shows that it has \emph{Universal Turing Machine} expressivity \cite{MuggletonLT15}. 
Accordingly, considering only hypotheses from the class $H^2_m$ does not constitute a severe restriction.

On the other hand, the set of hypotheses that can be learned by the encodings based on forward-
chained meta-rules and extensional BK is restricted compared to the solutions that can be obtained
by using the general encoding. In particular, as e.g.\ the meta-rule $P(x,y) \leftarrow Q(y,x)$ is not forward-chained, it is not possible to learn the inverse of binary predicates from the BK. For instance,
when a predicate $move\_right$ is contained in the BK, it is not possible to learn a predicate
$move\_left$. In this case, the inverses of binary predicates could be added to the BK explicitly.
Furthermore, the restriction to extensional BK prevents dependencies of the BK on the induced hypothesis, i.e.\
predicates in the BK cannot be defined in terms of predicates defined by a solution.
However, the majority of MIL-problems considered in the literature are forward-chained, and they do not
employ BK that depends on the respective hypothesis as usually only invented predicates are used
for rule heads in a hypothesis.

Intuitively, forward-chained meta-rules are natural for applications where binary examples
represent a mapping from their first to their second argument, e.g.\ of an initial state to a goal state
in a planning scenario or a string transformation, and where sequences of operations need to be
applied to obtain the second from the first argument. Many MIL-problems resulting from practical applications fall into this class.
Previous applications of MIL have been mainly considered in three different areas:
\emph{Robot Planning}, e.g.\ by Cropper and Muggleton \citeyear{CropperM14,CropperM15,CropperM16}; \emph{String/Language Processing}, e.g.\ by \citeN{LinDETM14} and by \citeN{CropperTM15}; and \emph{Computer Vision}, e.g.\ by \citeN{DaiMWTZ17}.
We found that most of the
MIL-problems considered in the first two areas
solely employ forward-chained
rules and extensional \BK, or in some cases can easily be transformed to forward-chained rules.
However, there are also some applications of MIL in the literature where a mapping to forward-chained
meta-rules is not (easily) possible \cite{Tamaddoni-Nezhad14,FarquharGCMB15,DaiMWTZ17}.

\leanparagraph{State Abstraction for Nondeterministic Planning Problems}
With respect to the degree of nondeterminism of the planning problems associated with a forward-chained MIL-problem, we can distinguish two factors that impact the size of the search space.
First, many different (potentially nondeterministic) actions may be applicable in the different states of a planning problem while there are only few valid plans. Second, there may also be many different action sequences constituting solutions to the respective planning problem. 

In the first case, the size of the search space 
can be reduced compared to Metagol by precomputing correct plans in the
encoding $\Pi_{sa}(\milp)$, which also reduces the size of the grounding. While Metagol
generates meta-substitutions based on all applicable actions, $\Pi_{sa}(\milp)$ only considers actions and states that are part of a correct plan.
In the second case, the size of the search space generated by $\Pi_{sa}(\milp)$ is closer to the size of the search space explored by Metagol because all possible plans need to be computed and considered
during induction. This is necessary since it cannot be decided beforehand which selection of plans allows for a
minimal solution wrt. the number of rules.
Note that, for the same reason, the search space of Metagol also must contain all possible plans wrt.\ positive examples.

Accordingly, the effectiveness of the encoding $\Pi_{sa}(\milp)$ depends on a
tradeoff between the number of actions applicable to states and the number of plans that can be generated for positive examples, and it has an advantage when there are many possible actions but only few plans. Due to the grounding bottleneck, the encoding $\Pi_{sa}(\milp)$ is likely to be less efficient than Metagol when there are many possible
plans and according states that need to be imported. It is an open challenge to tackle problems of this type efficiently by using state abstraction.

As noted in Section 4, our approach could be extended by filtering techniques
to preselect plans by the external atom in order to avoid the import of all possible plans for positive examples.
At the same time, this would be difficult to realize in Metagol where planning and generalization are performed simultaneously.
For instance, the number of plans could be restricted
by analyzing them, and filtering those that are redundant for obtaining a minimal solution.
Furthermore, 
only a limited number of plans could be imported and the impact on the accuracy investigated to find a good balance between efficiency and compactness of the hypothesis.

\leanparagraph{State Abstraction and Cyclic Sequences}
When computing sequences that derive positive examples according to Definition~\ref{def:sequences},
cyclic sequences need to be avoided.
At the same time, cyclic sequences potentially allow to induce a smaller hypothesis for a given MIL-problem, such that part (ii) of Theorem 3 is restricted to acyclic sequences as well.
However, note that in general,
a shorter sequence that derives the second argument of a positive examples from its first argument is obtained by removing cycles.
Hence, in practice, the prevention of cyclic plans is often reasonable as, e.g.\ considering a robot planning scenario, 
one does not want the robot to loop many times between the same states.
Furthermore, one is usually interested
in learning a strategy that generalizes minimal (or at least reasonably short) plans.
Consequently, there is a tradeoff between the lengths of plans that are considered for learning a strategy, and the size of a hypothesis that generalizes them.
Potentially more compact hypotheses can be obtained by allowing cyclic plans, but infinite loops
must be prevented.

One
way to relax the acyclicity condition would be to allow for a fixed number of cycles in Definition~\ref{def:sequences}, which may enable the induction of a smaller hypothesis. To empirically investigate the effect
of allowing different numbers of cycles in sequences wrt.\ the accuracy that can be achieved is  subject of future work.
Moreover, the possibility of cyclic sequences also poses a problem for termination in Metagol, where a different approach is taken to avoid infinite loops. It relies on ordering constraints over predicate arguments of meta-rules wrt.\ a total ordering over terms, which constrain the hypothesis space as well. Similar ordering information could alternatively be employed in our approach to prevent the generation of infinitely many sequences.

\leanparagraph{Related Work}
As discussed in Section \ref{intro}, our approach is most closely 
related to
\cite{MuggletonLPT14} 
which also applies ASP 
to MIL.
However, 
the ASP-encoding there is tailored to the induction of grammars, and
grounding issues or modeling a procedural bias
are not considered.

In addition, several other ILP systems based on ASP have been developed,
e.g.\ \cite{Otero01,Ray09,LawRB14},
which also mainly rely on an ASP-solver for computing solutions.
Different to our approach, \emph{default negation} is allowed in
the \BK\ and hypotheses, and induced programs
are interpreted under the \emph{stable model semantics}.
Moreover, examples are partial interpretations in the approach by \citeN{LawRB14}.
The declarative bias is defined by \emph{mode declarations} instead
of meta-rules in
the mentioned approaches, which enables a more fine-grained specification
of the hypothesis space; but, to the best of our knowledge,
none of them models a procedural bias wrt.\ rule introduction in ASP itself.
The \emph{XHAIL} system bounds the search space by splitting the learning
process into phases, where a \emph{Kernel set} of ground rules is computed
deductively and generalized in an \emph{induction phase}. However,
in contrast to the integration of object-level deduction and meta-level induction
in our encoding,
the phases are executed sequentially.

Compared to
ASP-based systems, the MIL-approach has the advantage
that meta-rules effectively limit the search space and, in
particular, 
can guide the process of predicate invention,
which is 
regarded 
as a very hard problem due to its high combinatorial
complexity \cite{DietterichDGMT08}.  In addition, intensional
\BK\ that manipulates complex terms is difficult to
integrate in ASP, while the query-driven
procedure exploited by Metagol is
well-suited for this.

\section{Conclusion}
We presented a general \hex-encoding
for solving MIL-problems that interacts with the \BK\ via
external atoms and restricts the search space by interleaving
derivations on the object and meta level. In addition, we introduced
modifications of
the encoding for certain types of MIL-problems and a
state abstraction technique to mitigate grounding issues
that are hard to tackle otherwise.

Our approach combines several advantages of Metagol and ASP-based approaches, and it
is very flexible as it allows to plug in arbitrary (monotonic) theories
as \BK.
Moreover, our encodings can easily be adjusted, e.g.\ by adding further 
constraints to limit the import of \BK.
For instance, we also tried to delay the import of \BK\
by restricting the initial import to chains of a limited length.
This resulted in a
significant 
speed-up for many MIL-problems, but minimality
of solutions
is not guaranteed.
Nevertheless, in our tests, solutions that were not considerably
larger  
than solutions of other instances
could be obtained 
instead of timeouts. 
To investigate how this and similar modifications affect the accuracy
wrt.\ a test data set remains future work.

The potential of an ASP-based approach for MIL is supported by our empirical evaluation;
and our techniques could also be exploited in future implementations.
Here, we use the \hex-formalism
because it is very convenient for prototype implementations. 
Other formalisms 
could be used as well,
e.g.\ the theory interface of \emph{Clingo~5} \cite{GebserKKOSW16}, 
which would potentially
increase performance.
In particular, employing optimization of \emph{weak constraints} is expected to
 be beneficial for efficiency as the solver needs to be
restarted many times during iterative deepening.

\section*{Acknowledgements}
This research has been supported by the Austrian Science
    Fund (FWF) projects P27730 and W1255-N23. Tobias Kaminski was supported by the NII International Internship Program. 
We thank the reviewers for their efforts and constructive suggestions, and Andrew Cropper
for helpful comments.

\bibliographystyle{acmtrans}
\bibliography{new_tlp2egui}

\newpage

\appendix

\section{Proof Sketches}

\begin{proof}[Proof Sketch for Theorem 1]
(i) The program $\mathcal{H}$ must be a solution for $\milp$ according to
Definition \ref{milproblem} because the
constraints in item (4) of Definition~\ref{hexmilenc} ensure that every
positive example $e^+ \in E^+$ is derivable by rules generated by the \BK\ imported in item (2), the rules generated by item (3b)
and meta-substitutions in $S$ obtained from guesses added by item (3a),
and that no negative example $e^- \in E^-$ is derivable.
In addition, atoms not imported from the \BK\ by item (2) are not relevant for deriving examples
according to Definition \ref{bkatoms}
as they are not entailed by $B \cup \mathcal{H}$.
Moreover, the facts generated by item (1) are only used in the positive bodies of
guessing rules generated by item (3a) such that they only constrain the
guesses for meta-substitutions. \\[0.15cm]
(ii) 
We know that $\mathcal{H}$ is a solution for $\milp$ s.t.\ 
all meta-substitutions in $\mathcal{H}$ satisfy the respective ordering
constraints and are productive.
Let $I$ be the interpretation containing
all atoms corresponding to facts generated by item (1) of Definition~\ref{hexmilenc}
wrt.\ $\milp$,
the atoms $unary(p,a)$ and $deduced(p,a,b)$ for all $p(a)$ and $p(a,b)$, resp., s.t.\
$B \cup \mathcal{H} \models p(a)$ and $B \cup \mathcal{H} \models p(a,b)$,
and
the atom
$meta(R_{id},p,q_1,...,q_k,r_1,...,r_n)$
for every rule $p(x,y) \leftarrow
q_1(x_1,y_1),...,q_k(x_k,y_k),r_1(z_1),...,$ $r_n(z_n) \in \mathcal{H}$
that is a meta-substitution of the meta-rule $R$.
Furthermore, let $I'$ be the interpretation containing all atoms
$n\_meta(R_{id},p,q_1,...,q_k,r_1,...,r_n)$ that occur in a ground
rule obtained from a rule generated by item (3b) whose body is satisfied by
$I$, s.t.\ $meta(R_{id},p,q_1,...,q_k,r_1,...,r_n) \not\in I$. It can be shown that $I \cup I'$ is an answer set of $\Pi(\milp)$ s.t.\
$\mathcal{H}$ is the logic program induced by $S$.
\end{proof}

\begin{proof}[Proof Sketch for Theorem 2]
(i) Compared to the general \hex-MIL-encoding of Definition~\ref{hexmilenc},
only item (2) is changed by the encoding $\Pi_f(\milp)$ such that the import
of \BK\ is guarded by the predicate $s$. As before,
item (4) ensures that every positive example $e^+ \in E^+$ is derivable.
It is only left to show that if a negative example $e^- \in E^-$ is entailed
by $B \cup \mathcal{H}$, then it can also be derived wrt.\ the \BK\ imported
via items (f1) and (f2) of Definition~\ref{fcenc}. Since all meta-rules
are assumed to be forward-chained, only meta-substitutions
of the form $ p(z_0,z_k) \leftarrow p(x,y),\dotsc,p_1(x_1,y_1),\dotsc,p_k(x_k,y_k),r_1(x_1),\dotsc,r_l(x_l)$ are usable for deriving examples
in which $x$ 
 is connected to $y$ by a chain of atoms $p_i(x_{i},y_i)$ in the body, where $y_i = x_{i+1}$, for $1 \leq i \leq k-1$, $x = x_1$ and $y = y_k$.
Furthermore, (f2) imports every binary atom in the \BK\ where the first argument
already occurs as first argument in an example or as second argument in an atom previously imported
from the \BK, due to items (f3) and (f4).

Similarly, all unary atoms in the \BK\ are imported by (f1) where the single argument occurs in a binary atom
from the \BK\ that has already been imported.
Hence, all \BK\ that is relevant for derivations by means of meta-substitutions wrt.\ forward-chained meta-rules is imported, and the second constraint of item (4) is violated in case
a negative example is entailed
by $B \cup \mathcal{H}$.\\[0.15cm]
(ii) Every minimal solution $\mathcal{H}$ for $\milp$ contains only productive
rules as defined right before Theorem 1 in Section \ref{sec:enc} because rules which are not
productive are not necessary for deriving a positive example.
Since we only consider forward-chained meta-rules, an answer set $S$ of $\Pi_f(\milp)$ such that $\mathcal{H}$ is the logic program induced by
$S$ only needs
to contain those binary atoms from the \BK\ that occur in a chain that connects
the first argument of each positive example $e^+ \in E^+$ to its second argument
because only those atoms are necessary for ensuring that each rule in $\mathcal{H}$ is productive.
Now, answer sets of $\Pi_f(\mathcal{M})$ are modulo the guess in (3a)
least models that can be constructed bottom up incrementally in a
fixpoint iteration, and that contain the atoms they logically entail.
Hence, all 
atoms in the corresponding chain are incrementally imported from the
\BK\ by the rules in items (f2) and (f4).
Accordingly, there is an answer set $S$ of $\Pi_f(\milp)$ s.t.\ the induced logic
program $\mathcal{H}$ wrt.\ $S$ is a minimal solution for
$\milp$.

Note that this suffices for finding minimal solutions in practice as our implementation
finds any productive solution that is encoded by $\Pi_f(\milp)$.
\end{proof}

\begin{proof}[Proof Sketch for Theorem 3]
This result can be shown similarly as the previous Theorem~2.
Each unary and binary atom introduced via the items (s1)
and (s2) of Definition~\ref{saencoding}, respectively, whose arguments occur
in an acyclic sequence of binary atoms from the \BK\ that connects
the first argument $a$ of each positive example $p(a,b) \in E^+$ to its second argument $b$,
can be mapped to exactly one unary and binary atom, resp., that is introduced
by items (f1) and (f2) of Definition~\ref{fcenc}.  
In this regard, the only difference
is that object-level constants in (f1) and (f2) are replaced by abstract states
of the form $(e^+_{id},seq_{id},i)$
according to Definition \ref{abstractimport} in (s1) and (s2).
Furthermore,  all acyclic sequences representing a possible chain that connects
the first argument of each positive example to its second argument are imported
by the external atom in item $(s4)$.

Then, the only essential remaining differences between $\Pi_f(\milp)$ and $\Pi_{sa}(\milp)$ consist in the fact
that sequences that are modeled by a solution and correspond to derivations of positive examples are guessed in item~(s4), and that instead of the second constraint from item~(4) of Definition~\ref{hexmilenc}, the derivability of negative examples is checked by means of the 
external atom in item~(s7).
However, a minimal solution for $\milp$ needs to model at least one sequence for deriving
each positive example, which is ensured jointly by items~(s3), (s5)
and (s6). 
Moreover, no minimal solution w.r.t.\ the restriction to acyclic sequences of Part (ii) of Theorem~3 is lost by excluding cyclic sequences in
Definition~\ref{def:sequences}.
Finally, item~(s7)  removes like the second constraint from item~(4)
all hypotheses that entail a negative example. 
\end{proof}

\section{Benchmark Encodings and Sample Instance}
To illustrate the concrete encodings and the instances employed for the empirical evaluation
in Section~5, we present the encodings
$\Pi_f(\milp)$ and $\Pi_{sa}(\milp)$ as well as the input to Metagol used for the \emph{Robot Waiter Strategies} benchmark (B3). A sample instance and a corresponding solution of benchmark (B3) can be
found at the end of this section.

Moreover, the encodings of all benchmark problems used in Section~5 and
all instances used in the experiments are available at http://www.kr.tuwien.ac.at/research/projects/inthex/hexmil/.

\subsection{Forward-Chained \hex-MIL-Encoding}
\small
\begin{verbatim}
binary(pour_tea,X,Y) :- &pour_tea[X](Y), state(X).
binary(pour_coffee,X,Y) :- &pour_coffee[X](Y), state(X).
binary(move_right,X,Y) :- &move_right[X](Y), state(X).

unary(wants_tea,X) :- &wants_tea[X](), state(X).
unary(wants_coffee,X) :- &wants_coffee[X](), state(X).
unary(at_end,X) :- &at_end[X](), state(X).

order(X,Y) :- skolem(X), binary(Y,_,_).
order(X,Y) :- pos_ex(X,_,_), binary(Y,_,_).
order(X,Y) :- pos_ex(X,_,_), skolem(Y).
order(X,Y) :- skolem(X), skolem(Y), X < Y.

{meta(precon,P1,P2,P3)} :- order(P1,P3), unary(P2,X), deduced(P3,X,Y).
{meta(postcon,P1,P2,P3)} :- order(P1,P2), deduced(P2,X,Y), unary(P3,Y).
{meta(chain,P1,P2,P3)} :- order(P1,P2), order(P1,P3), deduced(P2,X,Z),
                                                          deduced(P3,Z,Y).
{meta(tailrec,P1,P2,n)} :- order(P1,P2), deduced(P2,X,Z), deduced(P1,Z,Y).

deduced(P1,X,Y) :- meta(precon,P1,P2,P3), unary(P2,X), deduced(P3,X,Y).
deduced(P1,X,Y) :- meta(postcon,P1,P2,P3), deduced(P2,X,Y), unary(P3,Y).
deduced(P1,X,Y) :- meta(chain,P1,P2,P3), deduced(P2,X,Z), deduced(P3,Z,Y).
deduced(P1,X,Y) :- meta(tailrec,P1,P2,n), deduced(P2,X,Z), deduced(P1,Z,Y).

state(X) :- pos_ex(_,X,_).
state(Y) :- deduced(_,_,Y).

deduced(P,X,Y) :- binary(P,X,Y).

:- pos_ex(P,X,Y), not deduced(P,X,Y).
:- pos_ex(P,X,Y1), deduced(P,X,Y2), Y1 != Y2.

:- #count{ M,P1,P2,P3 : meta(M,P1,P2,P3) } != N, size(N).
\end{verbatim}
\normalsize

\subsection{State Abstraction \hex-MIL-Encoding}
Note that even though the syntax of the external atom used for importing binary and unary \BK\ as well
as the positive examples in the encoding below differs from the external atoms used in Definition~\ref{saencoding},
identical extensions are imported for the atoms $binary$, $unary$ and $pos$ as described in Section~4.
Hence, the encoding is equivalent to the encoding of Definition~\ref{saencoding}.

\small
\begin{Verbatim}[commandchars=\\\{\}]
binary(A,N1,N2) :- &abduceSequence[ID,ExStart,ExEnd](X,N1,N2,A),
                              pos_ex(ID,_,ExStart,ExEnd), X = seq.
unary(A,N) :- &abduceSequence[ID,ExStart,ExEnd](X,N,N,A),
                              pos_ex(ID,_,ExStart,ExEnd), X = check.
pos(ID,A,N1,N2) v n_pos(ID,A,N1,N2) :- 
                              &abduceSequence[ID,ExStart,ExEnd](X,N1,N2,A),
                              pos_ex(ID,_,ExStart,ExEnd), X = goal.

:- &failNeg[meta,pos_ex]().

pos1(ID) :- pos(ID,_,_,_).
:- pos_ex(ID,_,_,_), not pos1(ID).

order(X,Y) :- skolem(X), binary(Y,_,_).
order(X,Y) :- pos(X,_,_), binary(Y,_,_).
order(X,Y) :- pos(X,_,_), skolem(Y).
order(X,Y) :- skolem(X), skolem(Y), X < Y.

{meta(precon,P1,P2,P3)} :- order(P1,P3), unary(P2,X), deduced(P3,X,Y).
{meta(postcon,P1,P2,P3)} :- order(P1,P2), deduced(P2,X,Y), unary(P3,Y).
{meta(chain,P1,P2,P3)} :- order(P1,P2), order(P1,P3), deduced(P2,X,Z),
                                                          deduced(P3,Z,Y).
{meta(tailrec,P1,P2,n)} :- order(P1,P2), deduced(P2,X,Z), deduced(P1,Z,Y).

deduced(P1,X,Y) :- meta(precon,P1,P2,P3), unary(P2,X), deduced(P3,X,Y).
deduced(P1,X,Y) :- meta(postcon,P1,P2,P3), deduced(P2,X,Y), unary(P3,Y).
deduced(P1,X,Y) :- meta(chain,P1,P2,P3), deduced(P2,X,Z), deduced(P3,Z,Y).
deduced(P1,X,Y) :- meta(tailrec,P1,P2,n), deduced(P2,X,Z), deduced(P1,Z,Y).

deduced(P,X,Y) :- binary(P,X,Y).

:- not deduced(P,X,Y), pos(_,P,X,Y).

:- #count{ M,P1,P2,P3 : meta(M,P1,P2,P3) } != N, size(N).
\end{Verbatim}
\normalsize

\subsection{Metagol Input Program}
\small
\begin{verbatim}
metagol:functional.

func_test(Atom,PS,G):-
  Atom = [P,A,B],
  Actual = [P,A,Z],
\+ (metagol:prove_deduce([Actual],PS,G),Z \= B).

metarule(precon,[P,Q,R],([P,A,B]:-[[Q,A],[R,A,B]])).
metarule(postcon,[P,Q,R],([P,A,B]:-[[Q,A,B],[R,B]])).
metarule(chain,[P,Q,R],([P,A,B]:-[[Q,A,C],[R,C,B]])).
metarule(tailrec,[P,Q],([P,A,B]:-[[Q,A,C],[P,C,B]])).


prim(pour_tea/2).
prim(pour_coffee/2).
prim(move_right/2).

prim(wants_tea/1).
prim(wants_coffee/1).
prim(at_end/1).


a :-
  train_exs(Pos),
  Neg = [],
  learn(Pos,Neg).


pour_tea([robot_pos(X),end(Y),places([place(X,A,cup(up,empty))|R])],
                     [robot_pos(X),end(Y),places([place(X,A,cup(up,tea))|R])]).
pour_tea([robot_pos(X),end(Y),places([E|R1])],
                                       [robot_pos(X),end(Y),places([E|R2])]) :-
pour_tea([robot_pos(X),end(Y),places(R1)],[robot_pos(X),end(Y),places(R2)]).

pour_coffee([robot_pos(X),end(Y),places([place(X,A,cup(up,empty))|R])],
                  [robot_pos(X),end(Y),places([place(X,A,cup(up,coffee))|R])]).
pour_coffee([robot_pos(X),end(Y),places([E|R1])],
                                       [robot_pos(X),end(Y),places([E|R2])]) :-
pour_coffee([robot_pos(X),end(Y),places(R1)],[robot_pos(X),end(Y),places(R2)]).

move_right([robot_pos(X1),end(Y)|R],[robot_pos(X2),end(Y)|R]) :-
                                                          X1 < Y, X2 is X1 + 1.

wants_tea([robot_pos(X),end(_),places([place(X,tea,_)|_])]).
wants_tea([robot_pos(X),end(Y),places([_|R])]) :- 
                                    wants_tea([robot_pos(X),end(Y),places(R)]).

wants_coffee([robot_pos(X),end(_),places([place(X,coffee,_)|_])]).
wants_coffee([robot_pos(X),end(Y),places([_|R])]) :-
                                 wants_coffee([robot_pos(X),end(Y),places(R)]).

at_end([robot_pos(X),end(X)|_]).
\end{verbatim}
\normalsize

\subsection{Sample Instance and Solution}
\small
\begin{verbatim}
Instance:

pos_ex([robot_pos(1),end(3),places([place(1,coffee,cup(up,empty)),
place(2,coffee,cup(up,empty))])],
[robot_pos(3),end(3),places([place(1,coffee,cup(up,coffee)),
place(2,coffee,cup(up,coffee))])]).

pos_ex([robot_pos(1),end(6),places([place(1,coffee,cup(up,empty)),
place(2,coffee,cup(up,empty)),place(3,coffee,cup(up,empty)),
place(4,tea,cup(up,empty)),place(5,coffee,cup(up,empty))])],
[robot_pos(6),end(6),places([place(1,coffee,cup(up,coffee)),
place(2,coffee,cup(up,coffee)),place(3,coffee,cup(up,coffee)),
place(4,tea,cup(up,tea)),place(5,coffee,cup(up,coffee))])]).

Solution:

robot(A,B):-robot_1(A,B),at_end(B).
robot(A,B):-robot_1(A,C),robot(C,B).
robot_1(A,B):-robot_2(A,C),move_right(C,B).
robot_2(A,B):-wants_tea(A),pour_tea(A,B).
robot_2(A,B):-wants_coffee(A),pour_coffee(A,B).
\end{verbatim}
\normalsize

\label{lastpage}
\end{document}